\documentclass{article}

\usepackage{arxiv}

\usepackage[utf8]{inputenc} 
\usepackage[T1]{fontenc}    
\usepackage{hyperref}       
\usepackage{url}            
\usepackage{booktabs}       
\usepackage{amsfonts}       
\usepackage{nicefrac}       
\usepackage{microtype}      
\usepackage{lipsum}
\usepackage{cite}
\usepackage[pdftex]{graphicx} 
\usepackage{amsmath}
\usepackage{graphicx}
\usepackage[section]{placeins}
\usepackage{multirow}
\usepackage{multicol}
\usepackage{algorithm}
\usepackage{algorithmic}

\title{Downscaling Attack and Defense: Turning What You See Back Into What You Get}

\author{
  Andrew J. Lohn\\
  Center for Security and Emerging Technology, Georgetown University\\
  Washington, DC, USA \\
  \texttt{drew.lohn@georgetown.edu} \\
}

\begin{document}
\maketitle

\begin{abstract}
The resizing of images, which is typically a required part of preprocessing for computer vision systems, is vulnerable to attack. Images can be created such that the image is completely different at machine-vision scales than at other scales and the default settings for some common computer vision and machine learning systems are vulnerable. We show that defenses exist and are trivial to administer provided that defenders are aware of the threat. These attacks and defenses help to establish the role of input sanitization in machine learning.
\end{abstract}


\section{Introduction}
\label{sec:introduction}
\begin{figure}
  \centering
  \includegraphics[width=0.4\linewidth]{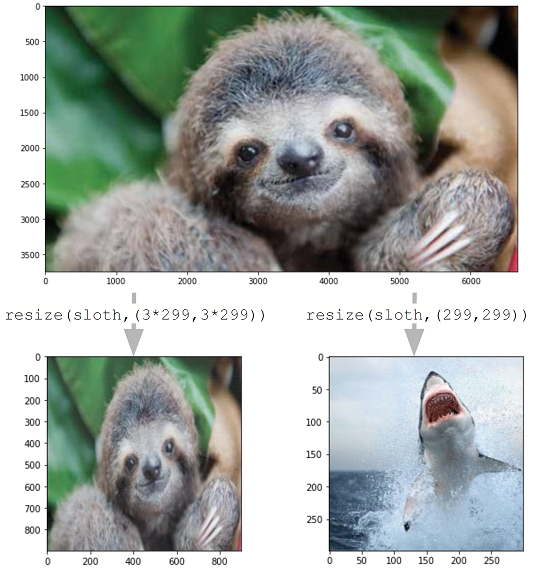}
  \caption{Resizing the same image to two different sizes can result in two completely different images.}
  \label{fig:basicPremise}
\end{figure}

Computer vision systems typically need to resize their inputs, and those resizing algorithms typically select only a small number of specific pixels from the original image. Those pixels can be specifically modified to contain the elements of a separate image that is completely different from the one that appears at other scales. This attack was first discovered in 2017 by Xiao et al. but without demonstrating defenses.\cite{firstDownscaling} The premise is shown in Figure \ref{fig:basicPremise}, where a high resolution sloth picture was manipulated so that it resizes to a shark at the 299x299 pixels shape used for the famous InceptionV3 classifier. The resizing algorithm used the default settings in TensorFlow: tf.image.resize(sloth,(299,299)).

Humans typically prefer to view images at a higher resolution than computer vision algorithms do, but also often at lower resolutions than what is provided by even commodity cameras. Using attacks like these, humans reviewing images at full or intermediate scales can be made to see benign images while the machine sees malicious inputs, or alternatively, humans may see restricted content that the machine marks as benign. For example, dissidents could use this technique to share or post images that appear contentious at human-scales but that automated censors view as completely benign.\cite{politicalMemes1,politicalMemes2} The outcome and applicability of these attacks is similar to that of adversarial examples but they are induced in a completely different way.\cite{intriguingProperties, explainingHarnessing, motivatingRulesOfGame}

Blending of images has been used in the design of triggers or backdoors\cite{targetedBackdoors, poisonFrogs} but those are unrelated to the present work. The images in downscaling attacks are combined in a different way that allows the images to be completely or nearly completely decoupled from each other. The threat model is also quite different. In the present attack, the attacker does not influence the training process to create a backdoor, they simply arrange the input so that different images are seen by the machine than by the human at another scale.

A useful way to think of these attacks is as artifacts of the rescaling step. Typically the goal of image processing is to minimize artifact generation, but here we are structuring the input image so as to maximize artifacts. Specifically, a second image is steganographically hidden so that it is revealed when rescaled to a specific size and shape. There are long histories in both steganography\cite{stegBook,stegPaper} and image processing.\cite{imageProcBook} This technique would not be prominent in either discipline. It is very ill-suited for steganography because the embedded information is much easier to detect and decode than many alternative steganographic methods. And image processing is almost exclusively directed at removing or reducing artifacts rather than increasing them.

This paper starts by introducing bilinear interpolation, which is a common default resizing algorithm though not the only vulnerable one.\cite{downscalingAlgorithms} We then describe the process of generating the attacks and defending against them. Then we discuss their implications for machine learning security.

\section{Bilinear Interpolation}
\label{sec:bilinearInterpolation}
\begin{figure}
  \centering
  \includegraphics[width=0.5\linewidth]{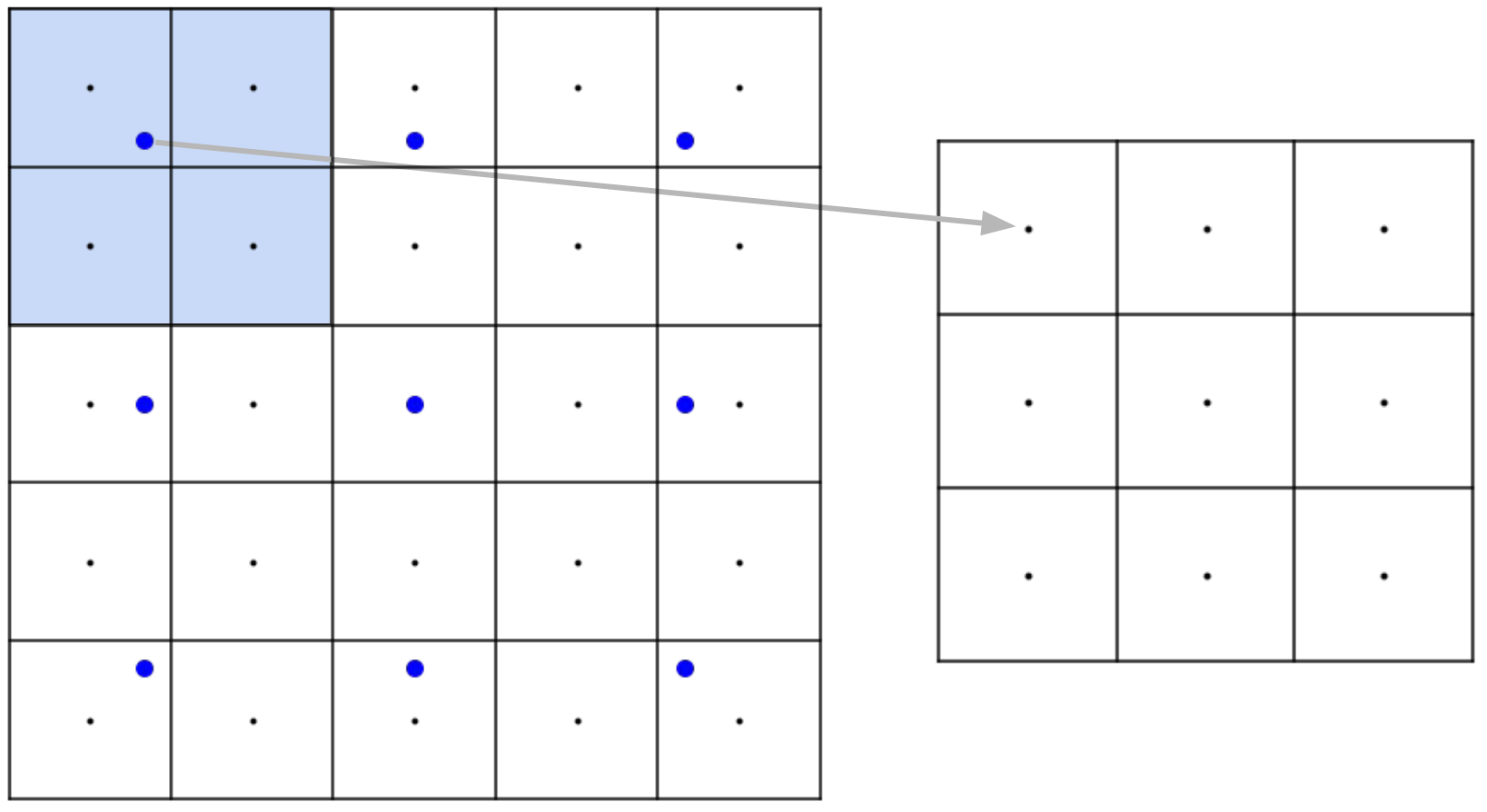}
  \caption{The four source image pixels from a 5x5 that contribute to the (1,1) pixel are highlighted.}
  \label{fig:interpLoc}
\end{figure}

A common default resizing method in image processing and machine learning libraries is bilinear interpolation. In it, each pixel in the resized image contains information from only four pixels in the source image. Those four pixels are the four diagonal neighbors in the source image that are closest to where the center of the corresponding pixel in the resized image would be if it was placed in the source image. The intensities of the four source image pixels are combined in two steps, first a weighted average in either the x or y directions, then a weighted average in the remaining direction. The locations of the four source image pixels are illustrated in Figure \ref{fig:interpLoc}.

Mathematically, let the source image have pixel locations \((n-0.5,m-0.5)\) for n and m ranging from \((1,1)\) to \((N,M)\) and pixel intensities \(f_{n,m}\) which may be vector-valued for color. And let the rescaled image have pixel locations \((i-0.5,j-0.5)\) ranging from \((1,1)\) to \((I,J)\) with pixel intensities \(g_{i,j}\). Then the continuous-valued location of the rescaled image pixel \((i,j)\) corresponds to the following location in the source image:

\begin{equation}
    (x,y) = ((i-0.5)\frac{N}{I},(j-0.5)\frac{M}{J})
\end{equation}

This continuous-valued location in the source image is then used to find the discrete locations of the four neighboring diagonal pixels with coordinates that are lower (\(l\)) and upper (\(u\)) with respect to the continuous-valued location. 

\begin{equation}
\begin{split}
    n_l = floor(x-0.5) \\
    m_l = floor(y-0.5) \\
    n_u = ceil(x-0.5) \\
    m_u = ceil(y-0.5)
\end{split}
\end{equation}

The resulting pixel intensity in the rescaled image is then calculated according to the following equation that interpolates the pixel values in both the x and y directions.

\begin{equation}
\label{bilinearBasics}
g_{i,j} = \frac{1}{(n_u - n_l)(m_u - m_l)}
\begin{pmatrix}n_u-x & x-n_l
\end{pmatrix}
\begin{pmatrix}f_{n_l,m_l} & f_{n_l,m_u} \cr f_{n_u,m_l} & f_{n_u,m_u} \cr
\end{pmatrix}
\begin{pmatrix}m_u-y \cr y-m_l \cr
\end{pmatrix}
\end{equation}

The weighting in equation \ref{bilinearBasics} may be important in more subtle attacks, but it does not play a role in any of the attacks discussed in this paper because we set all of the four diagonal pixel intensities to the same value. We simply put the full intensity of the desired rescaled pixel value in all four of the corresponding source image pixels.

\begin{figure}
  \centering
  \includegraphics[width=0.5\linewidth]{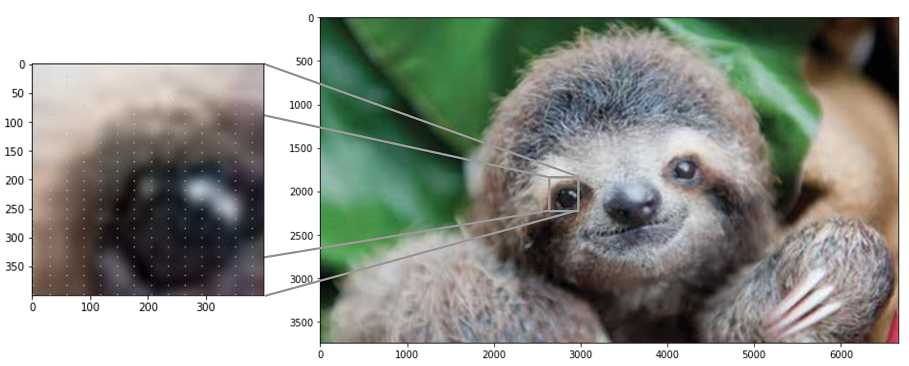}
  \caption{The repeating pattern of alterations is visible when zoomed in on the full-scale version.}
  \label{fig:zoomedIn}
\end{figure}

Figure \ref{fig:zoomedIn} shows the malicious pattern in a zoomed-in section of the full-scale image of the sloth from Figure \ref{fig:basicPremise}. Note that the pattern would be even less visible in a zoomed in cut of the rescaled sloth from Figure \ref{fig:basicPremise} because the intermediate downscaling selected sloth images pixels and little to no shark content.

\section{Generating Downscaling Attacks}
\label{sec:moireAttacks}
The process for generating downscaling attacks is simple. There is a big image like the sloth in the example we have adopted, there is a small image like the shark, and there is a combined image that has elements of the small image scattered throughout the big image. Creating the combined image is simply a matter of copying the right pixels of the small image to the right locations of the big image. The process is provided in pseudocode in Algorithm \ref{moireAlgorithm} using the same variable definitions as in Section \ref{sec:bilinearInterpolation}.

\begin{algorithm}
\begin{algorithmic}
\STATE $Combined\_Image \leftarrow Big\_Image$
\FOR{i <= I}
\FOR{j <= J}
\STATE $n_l \leftarrow floor(\frac{N(i+0.5)}{I}-0.5)$
\STATE $m_l \leftarrow floor(\frac{M(j+0.5)}{J}-0.5)$
\STATE $n_u \leftarrow ceil(\frac{N(i+0.5)}{I}-0.5)$
\STATE $m_u \leftarrow ceil(\frac{M(j+0.5)}{J}-0.5)$
\STATE
\STATE $intensity \leftarrow Small\_Image[i,j]$
\STATE $Combined\_Image[n_l,m_l] \leftarrow intensity$
\STATE $Combined\_Image[n_l,m_u] \leftarrow intensity$
\STATE $Combined\_Image[n_u,m_l] \leftarrow intensity$
\STATE $Combined\_Image[n_u,m_u] \leftarrow intensity$
\ENDFOR
\ENDFOR
\end{algorithmic}
\caption{Pseudocode outlining the process of generating downscaling attacks}
\label{moireAlgorithm}
\end{algorithm}

\begin{figure}
  \centering
  \includegraphics[width=0.5\linewidth]{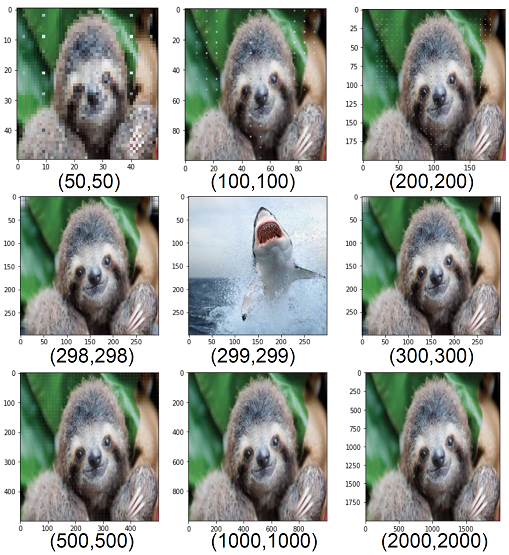}
  \caption{Elements of the shark are present at other scales but are only clearly reconstructed at the specified 299x299 scale.}
  \label{fig:manyScales}
\end{figure}

If there is a large difference in size between the small and big images then the altered pixels make up a small fraction of the combined image so that the effect can be minimal and difficult for humans to observe at full scale. As the difference in sizes grows, it also becomes increasingly rare for pieces of the smaller image to find their way into rescaled images at intermediate scales. As an illustration, the combined sloth-shark image (originally 3744x6660) was rescaled to various sizes as shown in Figure \ref{fig:manyScales}. At smaller scales, a substantial number of shark pixels are included in the downsampling and might or might not alert a human or secondary computer vision algorithm. At scales just one pixel above and below 299x299, the same is true but the alerting pixels are confined to the corners. At intermediate scales, larger than 299x299, the differences are even less perceptible and elements of the shark may not exist at all.

\begin{figure}
  \centering
  \includegraphics[width=0.4\linewidth]{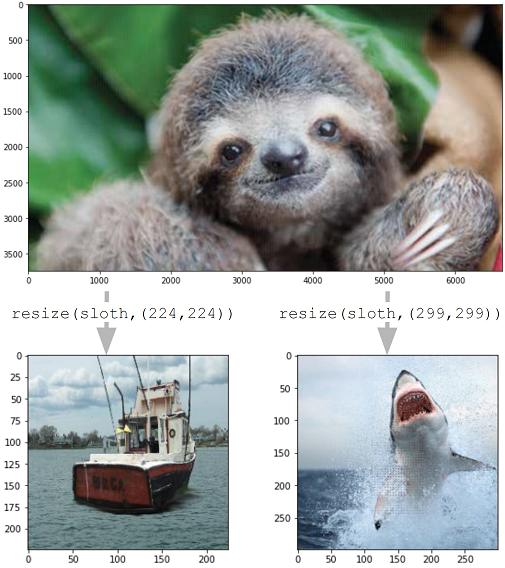}
  \caption{Downscaling attacks can be used to send different images to different classifiers, both of which are different from the large scale version.}
  \label{fig:threePics}
\end{figure}

It is also possible to embed a third image that is revealed at a third scale. Figure \ref{fig:threePics} shows the boat from the movie Jaws placed within the combined sloth-shark picture so that only the boat is seen at the 224x224, which is a common scale for other common image classifiers like VGG16. In generating the full-scale combined image, a few of the boat pixels land in the same spots as the shark pixels so overwrote them. That leads to some boat components in the belly of the shark that are visible on close inspection. Nonetheless, it is still clearly a sloth at large scale, a shark at Inception scale, and a boat at VGG scale, indicating that it is possible to create a single image that is viewed differently by humans and several different image classifiers.

\section{Defending Against Downscaling Attacks}
\label{sec:defenses}
Downscaling attacks are concerning because default settings in several cases are vulnerable, because they are very easy to generate, and because they are almost certain to be effective for any classifier that uses a vulnerable rescaling algorithm. But not all rescaling algorithms are vulnerable and it is fortunately a trivial matter both to detect these attacks and to harden rescaling algorithms. 

\begin{figure}
  \centering
  \includegraphics[width=0.7\linewidth]{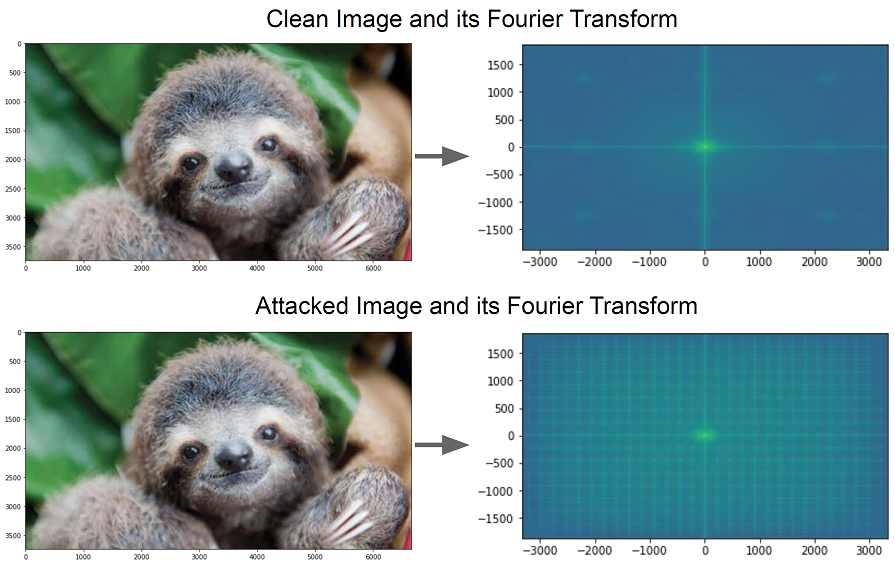}
  \caption{The attack is difficult to see in the images themselves but becomes obvious in their Fourier transforms.}
  \label{fig:fourier}
\end{figure}

When viewed at full-scale, these attacks include a regular repeating pattern, the frequency of which depends precisely on the size of the input image and of the rescaled image. So although the perturbations may be not be obvious, a human who knows to look for them can zoom in on the full-scale image and find the pattern. For more automated detection, the patterns become obvious if the image is transformed to the frequency domain as illustrated in Figure \ref{fig:fourier}. 

Detecting attacks may be less important than simply defanging them though, and defanging is even more straightforward. The attacks are possible because resizing algorithms only sample a few pixels from the source image for every pixel in the resized image. As described earlier, this sampling procedure can lead to artifacts in non-malicious cases as well, so techniques have already been developed to reduce the impact. One technique is to resize an image over a series of steps rather than all at once. This is the default approach used by the Python Pillow image processing library.  

\begin{figure}
  \centering
  \includegraphics[width=0.7\linewidth]{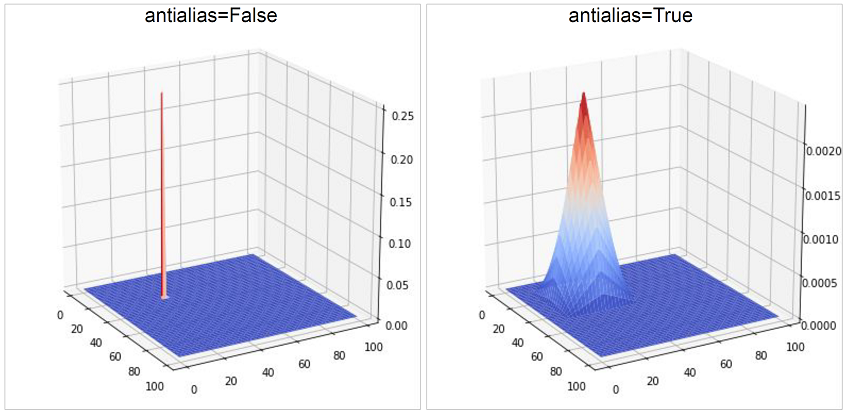}
  \caption{Contribution of each pixel in a 100x100 image to the (2,2) pixel when downsampled to a 5x5 image.}
  \label{fig:antialiasSampling}
\end{figure}

Another, perhaps more secure approach, samples a larger number of pixels in a one-step resizing in a form of antialiasing.\cite{antialiasingConvNet} Figure \ref{fig:antialiasSampling} shows the principle of antialiasing for scaling a 100x100 image to one that is only 5x5. The figure was generated by observing the intensity of the (2,2) pixel of the 5x5 image after rescaling from the 100x100 image with only one pixel turned on at a time. The idea is to see how much each pixel in the 100x100 image contributes to the (2,2) pixel in the 5x5 version. With antialias turned off, only four locations in the 100x100 image contribute, each contributing a quarter of its intensity. With antialias on, a much wider range of nearby pixels each contribute a much smaller fraction of their intensity. For a downscaling attack to succeed with antialiasing on would require extensive perturbations to the source image.

\begin{figure}
  \centering
  \includegraphics[width=0.9\linewidth]{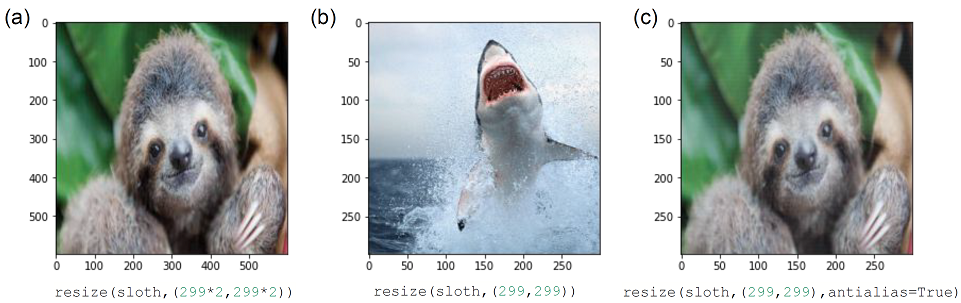}
  \caption{Moir\'e attacks are ineffective when antialiasing is turned on.}
  \label{fig:antialiasing}
\end{figure}

The practical effectiveness of antialiasing can be seen clearly in Figure \ref{fig:antialiasing}, where the TensorFlow default settings are used to generate a sloth at intermediate scales and a shark at the machine scale. When resized to the same machine scale with antialiasing on, the resulting image is once again a sloth. The antialiased sloth is slightly blurry, which is a standard artifact caused by antialiasing, but it is certainly not a shark to a human observer.

To gain a sense for the effectiveness of the attack and defenses, we collected 1000 random images from ImageNet\cite{imagenet_cvpr09} and tested two classifiers: InceptionV3\cite{inception} and VGG16.\cite{vgg} Inception used inputs scaled to 299x299 and VGG used 224x224. The inputs were scaled using the TensorFlow.image.resize() method, using both the default settings and again with antialiasing enabled. VGG requires additional preprocessing steps after rescaling for which we used the Keras preprocess\_input method with default settings. For combining the images, we selected randomly from the thousand images such that all thousand are included once as the big image and all thousand are included once as the small image with no repeats and none of the big images has itself as the small image. The combination process was conducted by resizing one randomly selected image to 2000x2000 (big image) and another other to 299x299 (small image) then combining them using Algorithm \ref{moireAlgorithm}.

Each input to the classifier has two images so also has two sets of labels to evaluate, one for the big images and one for the small images. Before evaluating the combined images however, we establish a baseline using clean images that do not have small images embedded in them. Across the top row of Table \ref{classiferResults}, the top-1 and top-5 accuracies match the reported accuracies for both Inception and VGG. The second row is also as expected: none of the small images were embedded, so the small labels should only be correct at about the one in a thousand and five in a thousand odds of random chance for top-1 and top-5 respectively.

The attacked images show a complete reversal. Inception, which uses the 299x299 scale that the attacks were designed for, sees the big images at only random chance and the embedded images at the full expected accuracy of Inception. VGG, operating at 224x224, on the other hand only sees the embedded images at random chance so is not susceptible to the same attack. It does have a significant drop in performance in observing the big images though. The VGG performance can be understood by revisiting Figure \ref{fig:manyScales} which showed the sloth example at many different scales. The shark was only clear at exactly the 299x299 scales so it is not surprising that the 224x224 scale of VGG results in only random chance for the small image labels. Degraded performance for the big image labels is also unsurprising because elements of the small image are likely to show up scattered around the rescaled image so, although it still mostly looks like the big image, it will be somewhat degraded, leading to decreased classifier accuracy as observed. It would be possible to tune the process to minimize the loss of accuracy at the 224x224 scale but we made no attempt to do so in this work.

We tested the effectiveness of defenses by turning on antialiasing during the downscaling process. Both of the classifiers took a small hit in performance even on clean images that had no small image embedded in them. The accuracy loss was about 2 percent and is likely due to the blurring effect caused by antialiasing. Rescaling with antialiasing can also be orders of magnitude slower for large images. For some, these may not be acceptable losses in performance. For others they may be welcome sacrifices for increased security. That security is substantial. With antialiasing on, the embedded images in the downscaling attacks were seen only at levels of random chance and the big images were seen in almost all cases. The big image accuracy for defended attacks was only a couple percent lower than for the clean images that were not attacked at all. The additional loss in accuracy as compared to the antialiased clean images is likely the result of the almost imperceptible small image pixels being blurred into the big image by the antialiasing procedure.

\begin{table}
\centering
    \begin{tabular}{|c|c|c|c|c|c|}
        \hline
         Images & Labels & Inception Top-1 & Inception Top-5 & VGG Top-1 & VGG Top-5 \\ \hline
         \multirow{2}{*}{Clean} & Big & 0.869 & 0.975 & 0.713 & 0.884 \\ \cline{2-6}
         & Small & 0.002 & 0.005 & 0.001 & 0.004 \\ \hline
         \multirow{2}{*}{Attacked} & Big & 0.002 & 0.003 & 0.343 & 0.593 \\ \cline{2-6}
         & Small & 0.866 & 0.975 & 0.000 & 0.004 \\ \hline
         \multirow{2}{*}{Antialiased Clean} & Big & 0.850 & 0.971 & 0.690 & 0.867 \\ \cline{2-6}
         & Small & 0.002 & 0.005 & 0.001 & 0.003 \\ \hline
         \multirow{2}{*}{Antialiased Attacked} & Big & 0.821 & 0.955 & 0.677 & 0.855 \\ \cline{2-6}
         & Small & 0.003 & 0.006 & 0.001 & 0.002 \\ \hline
    \end{tabular}
    \caption{The attacks are nearly perfectly effective against the Inception classifier they were designed for and almost perfectly ineffective in the presence of antialiasing. For VGG, which uses a different scale for its inputs, the attacks almost never succeed at selecting the embedded image but performance at identifying the bigger image is degraded.}
    \label{classiferResults}
\end{table}

\section{Discussion}
\label{sec:discussion}
Downscaling attacks are nearly one hundred percent effective against vulnerable preprocessing procedures. Knowing that these attacks exist though, they are trivial to detect and they are trivial to defend against with only limited loss in accuracy. Unfortunately, some of the most popular preprocessing procedures are currently vulnerable by default, and some of those that are not can be made vulnerable by an ill-intentioned designer just as easily as they can be hardened. TensorFlow and OpenCV are susceptible by default. Skimage, a package for sklearn, has anti-aliasing on by default but can be made susceptible by setting it to False. Pillow uses a multi-step downscaling so attacking it is harder. PyTorch uses Pillow in torchvision so is less vulnerable, but an alternative torchvision has been written to use the faster OpenCV instead, which is susceptible by default.

The practice of restructuring or filtering inputs for security is relatively new for machine learning but is standard for conventional digital security. SQL injection and cross-site scripting are among the most basic techniques in a conventional hacker's toolkit. They work by inputting commands to be executed rather than just data to be processed, and they have caused vast numbers of breaches.\cite{akamaiWebAttacks} They still cause high profile breaches today, but it has become harder for attackers because defenders can process out malicious inputs in what is known as input sanitization.\cite{sanitization} Machine learning security is still at the stage of learning what attacks are in the realm of the possible and what it even means to sanitize inputs. For example, antialiasing,\cite{featureSqueezing, sparseCoding} or alternatively, adding noise to inputs,\cite{advNoise1,advNoise2} has been shown to be a partially effective protection against some adversarial examples. As shown in this paper, ensuring that antialiasing is on can be useful for preventing attacks that use artifacts of the rescaling process. It is likely that more attacks on the inputs are left to be discovered and that both the concept and practice of input sanitization for machine learning has more evolution to do.

\section{Conclusions}
\label{sec:conclusions}
Downscaling attacks, that take advantage of the repeating patterns inherent in downsampling an image during resizing, are a vulnerability for image classifiers. An image can be embedded within another image such that, when viewed at two different scales, the two appear completely different. That makes it possible for attackers to completely fool machine vision systems into arbitrary erroneous conclusions. But these attacks can be easily detected and easily defeated, and only some of the preprocessing methods of popular machine learning platforms are vulnerable by default. Sanitizing inputs by simply enabling antialiasing comes at a performance cost of only a couple percent while completely eliminating the threat, at least as posed in the current work. Knowing that these attacks exist is therefore much more than half the battle.

\section*{Acknowledgement}
The author would like to thank Ben Buchanan, John Bansemer, Patrick Mickel, and Gavin Hartnett for useful discussions.

\bibliographystyle{unsrt}  
\bibliography{references}

\begin{thebibliography}{10}

\bibitem{firstDownscaling}
Qixue Xiao, Kang Li, Deyue Zhang, and Yier Jin.
\newblock Wolf in sheep's clothing - the downscaling attack against deep
  learning applications.
\newblock {\em arXiv:1712.07805v1}, 2017.

\bibitem{politicalMemes1}
An~Xiao Mina.
\newblock Batman, pandaman and the blind man: A case study in social change
  memes and internet censorship in china.
\newblock {\em Journal of Visual Culture}, 13:359--375, 2014.

\bibitem{politicalMemes2}
Fan Yang.
\newblock Rethinking china's internet censorship: The practice of recoding and
  the politics of visibility.
\newblock {\em New Media \& Society}, 18:1364--1381, 2016.

\bibitem{intriguingProperties}
Christian Szegedy, Wojciech Zaremba, Ilya Sutskever, Joan Bruna, Dumitru Erhan,
  Ian Goodfellow, and Rob Fergus.
\newblock Intriguing properties of neural networks.
\newblock {\em arXiv:1312.6199}, 2014.

\bibitem{explainingHarnessing}
Ian Goodfellow, Jonathon Shlens, and Christian Szegedy.
\newblock Explaining and harnessing adversarial examples.
\newblock In {\em International Conference on Learning Representations}, 2015.

\bibitem{motivatingRulesOfGame}
Justin Gilmer, Ryan~P. Adams, Ian Goodfellow, David Andersen, and George~E.
  Dahl.
\newblock Motivating the rules of the game for adversarial example research.
\newblock {\em arXiv:1807.06732}, 2018.

\bibitem{targetedBackdoors}
Xinyun Chen, Chang Liu, Bo~Li, Kimberly Lu, and Dawn Song.
\newblock Targeted backdoor attacks on deep learning systems using data
  poisoning.
\newblock {\em arXiv:1712.05526}, 2017.

\bibitem{poisonFrogs}
Ali Shafahi, W.~Ronny Huang, Mahyar Najibi, Octavian Suciu, Christoph Studer,
  Tudor Dumitras, and Tom Goldstein.
\newblock Poison frogs! targeted clean-label poisoning attacks on neural
  networks.
\newblock {\em arXiv:1804.00792}, 2018.

\bibitem{stegBook}
Ingemar~J. Cox, Matthew~L. Miller, Jeffrey~A Bloom, Jessica Fridrich, and Ton
  Kalker.
\newblock {\em Digital Watermarking and Steganography}.
\newblock Morgan Kaufmann, 2 edition, 2008.

\bibitem{stegPaper}
Neil~F. Johnson and Suhil Jajodia.
\newblock Exploring steganography: Seeing the unseen.
\newblock {\em Computer}, 31:26--34, 1998.

\bibitem{imageProcBook}
Rafael Gonzalez and Richard Woods.
\newblock {\em Digital Image Processing}.
\newblock Pearson, 4 edition, 2018.

\bibitem{downscalingAlgorithms}
Yujia Liu, Weiming Zhang, and Nenghai Yu.
\newblock Query-free embedding attack against deep learning.
\newblock In {\em IEEE International Conference on Multimedia and Expo}, 2019.

\bibitem{antialiasingConvNet}
Xueyan Zou, Fanyi Xiao, Zhiding Yu, and Yong~Jae Lee.
\newblock Delving deeper into anti-aliasing in convnets.
\newblock {\em arXiv:2008.09604}, 2020.

\bibitem{imagenet_cvpr09}
J.~Deng, W.~Dong, R.~Socher, L.-J. Li, K.~Li, and L.~Fei-Fei.
\newblock {ImageNet: A Large-Scale Hierarchical Image Database}.
\newblock In {\em CVPR09}, 2009.

\bibitem{inception}
Christian Szegedy, Vincent Vanhoucke, Sergey Ioffe, Jonathon Shlens, and
  Zbigniew Wojna.
\newblock Rethinking the inception architecture for computer vision.
\newblock {\em arXiv:1512.00567}, 2015.

\bibitem{vgg}
Karen Simonyan and Andrew Zisserman.
\newblock Very deep convolutional networks for large-scale image recognition.
\newblock {\em arXiv:1409.1556}, 2014.

\bibitem{akamaiWebAttacks}
Monique Bonner.
\newblock Web attacks and gaming abuse.
\newblock Technical report, Akamai, 2019.

\bibitem{sanitization}
Lwin~Khin shar and Hee Beng~Kuan Tan.
\newblock Mining input sanitization patterns for predicting sql injection and
  cross site scripting vulnerabilities.
\newblock In {\em International Conference on Software Engineering}, 2012.

\bibitem{featureSqueezing}
Weilin Xu, David Evans, and Yanjun Qi.
\newblock Feature squeezing: Detecting adversarial examples in deep neural
  networks.
\newblock In {\em Network and Distributed Systems Security Symposium}, 2018.

\bibitem{sparseCoding}
Edward Kim, Jessica Yarnall, Priya Shah, and Garrett~T. Kenyon.
\newblock A neuromorphic sparse coding defense to adversarial images.
\newblock In {\em Proceedings of the International Conference on Neuromorphic
  Systems}, 2019.

\bibitem{advNoise1}
Shixiang Gu and Luca Rigazio.
\newblock Towards deep neural network architectures robust to adversarial
  examples.
\newblock In {\em International Conference on Learning Representations}, 2015.

\bibitem{advNoise2}
Bai Li, Changyou Chen, Wenlin Wang, and Lawrence Carin.
\newblock Certified adversarial robustness with additive noise.
\newblock In {\em Conference on Neural Information Processing Systems}, 2019.

\end{thebibliography}

\end{document}